\documentclass[prx,superscriptaddress,longbibliography,twocolumn]{revtex4-2}
\usepackage{bm}
\usepackage{amsmath, amsfonts}
\usepackage{amssymb}
\usepackage{times}
\usepackage{graphicx}
\usepackage{xcolor}
\usepackage{mathtools}
\usepackage{physics}
\usepackage{float}
\usepackage[colorlinks=true,allcolors=blue]{hyperref}
\usepackage[capitalise,nameinlink]{cleveref}
\usepackage{tikz}
\crefname{section}{Sec.}{Figs.}
\graphicspath{{Figures/}}
\begin{document}

\title{Revisiting the multi-mode rhombus circuit as a biased-noise qubit}

\author{Pablo Aramburu Sanchez}
\affiliation{Department of Physics, University of Colorado Boulder, Boulder CO 80309, USA}
\affiliation{Department of Electrical, Computer \& Energy Engineering, University of Colorado Boulder, Boulder, CO 80309, USA}

\author{Trevyn  F. Q. Larson}
\affiliation{Department of Physics, University of Colorado Boulder, Boulder CO 80309, USA}
\affiliation{National Institute of Standards and Technology, Boulder, Colorado 80305, USA}

\author{Anthony P. McFadden}
\affiliation{National Institute of Standards and Technology, Boulder, Colorado 80305, USA}

\author{Constantin Schrade}
\affiliation{Hearne Institute of Theoretical Physics, Department of Physics \& Astronomy, Louisiana State University, Baton Rouge LA 70803, USA}

\author{Joshua Combes}
\affiliation{School of Physics and School of Mathematics, University of Melbourne, Parkville, Victoria, 3010, Australia}

\author{Andr\'as Gyenis}
\email{andras.gyenis@colorado.edu}
\affiliation{Department of Electrical, Computer \& Energy Engineering, University of Colorado Boulder, Boulder, CO 80309, USA}
\affiliation{Department of Physics, University of Colorado Boulder, Boulder CO 80309, USA}

\begin{abstract}
    In this work, we revisit the idea of using an interferometer of pairs of Josephson junctions as a protected rhombus qubit. Unlike in the original proposal, where the qubit states are encoded into odd and even parity charge states, here, we intentionally alter the energy of one of the junctions to investigate the soft version of the rhombus qubit. This approach allows us to directly probe the qubit transitions over several GHz and reduce the potential drawbacks of the interferometer-based protection. Away from a half flux quantum external field, the large shunting capacitors of the circuit ensure localized qubit states in different phase valleys, leading to a biased-noise qubit. In the realized circuit, we measure an average $T_1\approx500\,\mu$s relaxation time in the biased-noise regime (with a Ramsey dephasing time of $T^{R}_\varphi\approx90\,$ns), while an average $T_1\approx27\,\mu$s relaxation time at frustration (with $T^{R}_\varphi\approx670\,$ns). Our loss analysis on this multi-mode circuit indicates that at low frequencies, flux noise and quasiparticle tunneling limit the relaxation times, pointing toward the presence of an optimal operating regime of around a few GHz. 
\end{abstract}

\maketitle 

\section{Introduction}

\begin{figure}[t]
\centering
\includegraphics[width = 8.6cm]{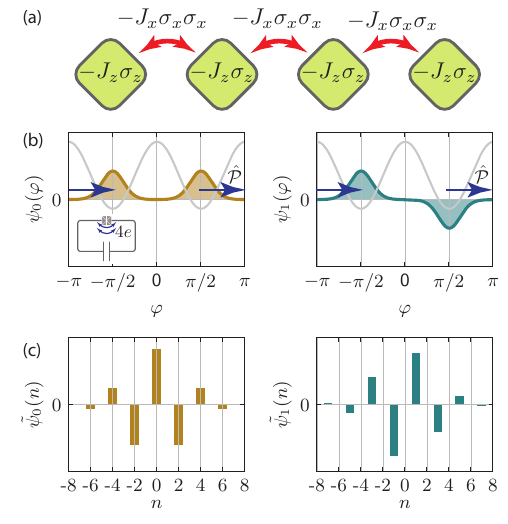}
\caption{\label{fig:fig1} \textbf{The rhombus qubit as the building element for a protected array qubit} (a) Schematics of a protected array qubit composed of single qubits (green rhombi) coupled through a strong longitudinal interaction (red arrows). Here $\sigma_z$ and $\sigma_x$ are the Pauli $z$ and $x$ operators, $2J_z$ is the qubit energy, $J_x$ is the coupling strength. (b), (c) The wavefunctions of the ground and excited states of the charge-parity qubit in phase and charge representations. The inset in b) shows the circuit diagram of the qubit, where the coherent tunneling of only $4e$ charges is allowed. Arrows indicate the effect of the charge-parity operator $\hat{\mathcal{P}}$.}
\end{figure}

Decoupling superconducting qubits from environmental noise has been a major goal since the inception of quantum circuits~\cite{Kitaev_2003, ioffe_environmentally_1999, Ioffe2002, PhysRevB.66.224503, PhysRevLett.90.107003, PhysRevB.71.024505}. Besides harnessing smart circuit designs to shield qubits from various errors~\cite{IoffeBDoucotLB2012, gyenis_moving_2021,PhysRevA.87.052306}, a possible route toward extending coherence times is to create larger \textit{array qubits} from lattices of faulty qubits~\cite{PhysRevLett.88.227005,gladchenko_superconducting_2009, PhysRevLett.112.167001, PRXQuantum.3.030303, PhysRevLett.131.150602}. The design of certain array qubits is motivated by well-studied condensed matter systems, such as the transverse-field Ising model, where single qubits are coupled through a strong interaction (Fig.~\ref{fig:fig1}a). This arrangement can lead to protection because when the coupling strength is much larger than the frequencies of the individual qubits, the splitting between the ground and the first excited states of the array becomes exponentially suppressed as the number of qubits is increased~\cite{PRXQuantum.3.030303}. 

Historically, one of the earliest proposals for a single qubit that could be used for such an Ising-type array qubit is the interference-based rhombus qubit~\cite{PhysRevB.63.174511, PhysRevLett.88.227005, gladchenko_superconducting_2009, PhysRevB.70.184519, PhysRevB.78.104504,PhysRevX.15.011021}. This circuit consists of four identical Josephson junctions in a loop, and has protected qubit states when biased at exactly half a flux quantum (Fig.~\ref{fig:fig1}b,c). However, when flux noise drives the qubit away from frustration, the protection vanishes, and the energy of the qubit becomes linearly sensitive to the magnetic field, which can be addressed by concatenating multiple rhombi (Fig.~\ref{fig:fig1}a). The rhombus qubit also served as one of the earliest realizations of protected qubits and linked to a plethora of new circuit designs, such as the two-Cooper-pair tunneling element~\cite{smith_superconducting_2020}, the kinetic interference cotunneling element~\cite{PhysRevX.12.021002},  the mirror-current qubit~\cite{kitaev2006protectedqubitbasedsuperconducting,PhysRevB.100.224507}, the $0-\pi$ qubit~\cite{PhysRevA.87.052306,PRXQuantum.2.010339}, the parity-protected superconductor-semiconductor qubit~\cite{PhysRevLett.125.056801, feldsteinbofill2026}, the bifluxon qubit~\cite{PRXQuantum.1.010307}, the non-degenerate noise-resilient qubit~\cite{hays2025nondegenerate}, and the $\cos2\varphi$ transmon~\cite{roverch2026}. 

Despite its considerable legacy, the study of various noise sources in a single rhombus has been lacking, which motivated us to revisit this old circuit.  While a perfect rhombus indeed possesses protection against dielectric-type relaxation processes, the interferometer-based layout and its low frequency expose the qubit to flux and quasiparticle noise. Thus, here, we modify its design such that the protection is more tailored towards realistic noise sources~\cite{messelot2026}. In our \textit{soft-rhombus} qubit, we intentionally introduce an asymmetry in the junctions that raises the energy of the qubit close to 100 MHz, while keeping the qubit charge-insensitive. Although this choice breaks a certain Cooper-pair number symmetry, it eventually leads to advantageous properties regarding the noise processes that matter at low frequencies. Furthermore, away from frustration, where protection arises from the qubit wavefunctions having a disjoint support, the device can be operated as a biased-noise qubit potentially with lower error correction thresholds when paired with tailored codes~\cite{doi10.1126,PhysRevLett.120.050505}.

\subsection{The ideal charge-parity protected qubit}

Before we turn to the rhombus qubit, we briefly review the ideal charge-parity protected qubit that the rhombus circuit can emulate. In the charge-parity protected qubit, a large capacitor shunts a hypothetical double-Cooper-pair tunneling element, which allows only pairs of Cooper pairs to tunnel through. The Hamiltonian in the charge basis is
\begin{equation}
     \hat{H}_\mathrm{CP} =4E_C\sum_n(n-n_g)^2|n\rangle\langle n| + \frac{E_2}{2}\sum_n\biggl(|n+2\rangle\langle n| + h.c. \biggl ), 
\end{equation}
where $E_C$ is the charging energy associated with the shunting capacitor, $E_2$ is the amplitude of the double-Cooper pair tunneling events, $n$ is the number of Cooper pairs across the junction, $n_g$ is the offset charge, and $|n\rangle$ is the eigenstate of the Cooper pair number operator $\hat{n}$ with the value of $n$. Equivalently, in the phase representation of the circuit Hamiltonian, the tunneling element has a $\cos(2\varphi)$ double-well potential, where $\varphi$ is the superconducting phase across the junction, such that
\begin{equation}
      \hat{H}_\mathrm{CP} =4E_C(-\mathrm{i}\partial_\varphi-n_g)^2 + E_2\cos 2\hat\varphi.
\end{equation}
In this qubit, the phase coordinate is $2\pi$-periodic, and the charge coordinate is discrete. We choose the convention where the potential energy term has a positive coefficient, so that the potential minima are located at $\varphi=\pm\pi/2$. The two lowest-lying levels define the computational qubit states, which are the superposition of even or odd charge states (Fig.~\ref{fig:fig1}c). 

The qubit states are eigenstates of the charge parity operator $\hat{\mathcal{P}}=e^{\mathrm{i}\pi \hat{n}}$. The charge parity symmetry ensures that local operators in charge space, such as the charge operator $\hat{n}$, can not introduce bit flip errors. From another perspective, the key to the protection is that the double-well potential minima are $\pi$ phase away, and the qubit wavefunctions in phase space are eigenstates of a $\pi$ phase shift operator. We emphasize that the qubit is not protected against processes where the noise couples to the $\cos(\hat\varphi)$ or $\sin(\hat\varphi)$ operators, because those operators describe hopping between neighboring charge states. This is crucial for the interferometer-based implementations, where flux noise can introduce coupling to these operators. Finally, we note that the charge parity symmetry alone is insufficient for phase-flip protection, but when a large shunting capacitance shunts the junctions, the dephasing rates due to charge noise can be strongly reduced. 

\subsection{Charge pairing through interference and the rhombus qubit}

\begin{figure}[t]
\centering
\includegraphics[width = 8.6cm]{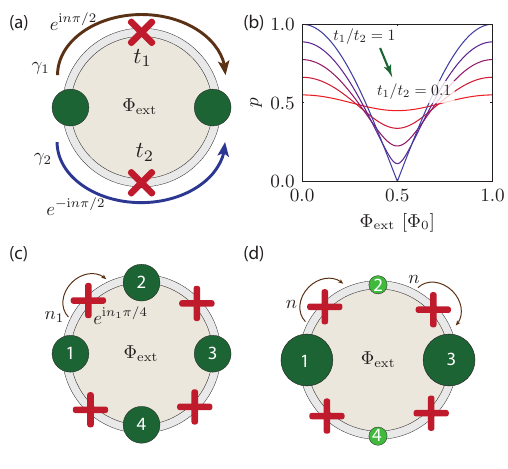}
\caption{\label{fig:fig2} \textbf{Interference as the principle behind the rhombus} (a) Interferometer loop enclosing a magnetic flux $\Phi_\mathrm{ext}$  and connecting two islands (green dots) through two tunnel barriers (red crosses). A charge can tunnel between the islands through two different paths and pick up phases $e^{\pm \mathrm{i}n\pi/2}$ when the loop is biased at frustration. (b) Single-Cooper-pair tunneling probability between the two islands as a function of external flux in the two-junction interferometer. The colors of the curves show the degree of the tunneling symmetry ($t_1/t_2$). Complete destructive interference arises when $t_1=t_2$. In the asymmetric case, the suppression of the transmission at $\Phi_0/2$ is the remnant of Cooper pair pairing, which is the principle behind the soft-rhombus qubit.  (c) Schematics of the rhombus, where different charges can tunnel between the islands, leading to various Aharonov-Bohm phase factors. For example, when $n_1$ charge tunnels between islands 1 and 2, the phase factor is $e^{\mathrm{i}n_1\pi/4}$ for the symmetric allocation of the vector potential. (d) The rhombus qubit, when two islands form a large capacitance across the circuit, and charging of the small islands is energetically unfavorable.}
\end{figure}

A possible route to physically realize the charge-parity protected qubit is to harness the Aharonov-Bohm interference. This effect is based on when $n$ Cooper pairs simultaneously travel along a path $\gamma$ with vector potential $\mathbf{A}$, their wavefunction acquires a phase shift of $\theta = 2\pi n {\int_{\gamma}\mathbf{A}d\mathbf{l}}/{\Phi_0}$, where $\Phi_0=h/2e$ is the flux quantum. For example, when Cooper pairs can traverse between two opposite points of a superconducting symmetric ring enclosing external flux $\Phi_\mathrm{ext}$, they can take a clockwise path $\gamma_1$ or the counterclockwise path $\gamma_2$, leading to a phase difference of $\Delta\theta= 2\pi n {\Phi_\mathrm{ext}} /{\Phi_0}$ (Fig.~\ref{fig:fig2}a). Thus, when $\Phi_\mathrm{ext} = \Phi_0/2$, the phase difference becomes $\Delta\theta = \pi n$, yielding constructive (destructive) interference for co-tunneling of an even (odd) number of Cooper pairs. Critically, this scheme works only when the tunneling amplitudes through the two arms are equal. As Fig.~\ref{fig:fig2}b shows, if the transmission coefficients are unbalanced ($t_1 \neq t_2$), the interference will not be fully constructive or destructive. 

Physically, one can implement such an interferometer by placing two Josephson junctions as barriers in the two arms--as long as the junctions support co-tunneling of Cooper pairs. In other words, the current-phase relation of the junctions needs to deviate from a pure sinusoidal form with higher harmonics present. This is the case for highly transmissive Josephson junctions, such as hybrid superconducting-semiconducting junctions~\cite{PhysRevLett.125.056801} or weak link junctions~\cite{likharev_superconducting_1979}. In contrast, for the widely used aluminum-oxide-based junctions, the co-tunneling of $n$ Cooper pairs is strongly suppressed~\cite{Willsch2024, kim_emergent_2025}, and the total Josephson energy diminishes in the interferometer at frustration. To circumvent this issue, the rhombus qubit was proposed (Fig.~\ref{fig:fig2}c), where two junctions relying on single-Cooper-pair tunneling are placed in series in each arm, which can mimic the behavior of junctions with higher harmonics~\cite{10.21468/SciPostPhys.15.5.204}.

Because the rhombus qubit has additional islands, its interference pattern is richer than that of a two-barrier interferometer. For example, the Aharonov-Bohm phase will depend on the number of charges tunneled between the various islands, leading to both fully and partially destructive interference effects (Fig.~\ref{fig:fig2}c). We note that this interference pattern becomes simpler when a large capacitor is placed across the device, for example, between islands 1 and 3 (Fig.~\ref{fig:fig2}d). In this case, the charging energies of the small islands 2 and 4 are large, making it energetically unfavorable to occupy these islands. Thus, we can recover the simpler case of a two-barrier interferometer. This capacitor configuration is similar to the setup used in transport measurements, when current leads are placed across the device, leading to double-Cooper pair transport~\cite{leblanc_gate-_2025,PhysRevB.78.104504, PhysRevX.15.011021,PhysRevLett.133.106001}. 

\begin{figure*}[t]
\centering
\includegraphics[width = \textwidth]{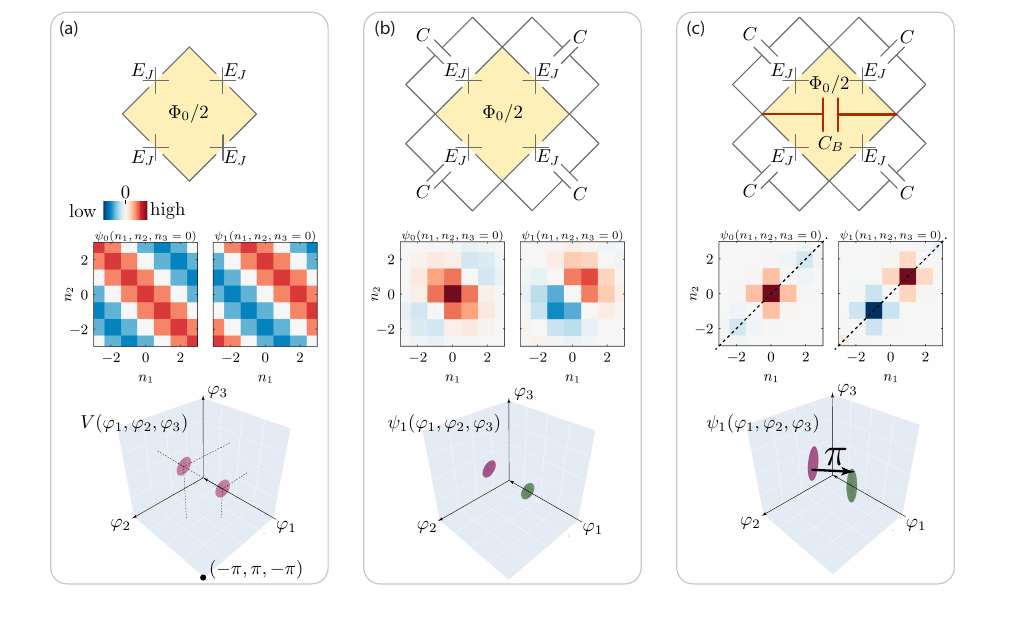}
\caption{\label{fig:fig3} \textbf{The potential and wavefunctions of the rhombus} (a) Theoretical limit, when the qubit phase wavefunctions are approximated by the symmetric and antisymmetric combinations of Dirac deltas located at the minima of the potential at $\Phi_\mathrm{ext}=\Phi_0/2$. Top panel: circuit schematics of the rhombus. Middle panel: a slice of the wavefunctions in charge representation showing a periodic interference-like pattern, where certain charge states are not allowed. Bottom panel: the surface plot of the potential with two minima at $(\varphi_1,\varphi_2,\varphi_3) = (\pm\pi/4,\pm\pi/4,\pm\pi/4)$. The dashed lines are guides to the eye for the locations of the potential minima. (b) The symmetrically shunted rhombus (top panel). The wavefunctions in charge space show a charge pattern similar to (a) with an additional decay function (middle panel). The surface plot of the wavefunction of the asymmetric excited state in phase space (bottom panel). (c) The rhombus qubit, when shunted by a large capacitor between two opposite islands (top panel). In charge space, along the $n_1=n_2$ direction (dashed lines), even and odd states are allowed for the ground and excited states (middle panel). In phase space, the wavefunction gets squeezed in such a way that the distance between the two lobes becomes approximately $\pi$ (bottom panel). In this sense, the rhombus qubit can approximate a $\cos2\varphi$ element between the two opposite nodes.}
\end{figure*}

\section{Results}

\subsection{Numerical simulations}

Now, we turn to the circuit quantization of the rhombus to obtain the energies and wavefunctions of the qubit (see Methods~\ref{app:cq} for details). The device has three degrees of freedom, and the Hamiltonian reads 
\begin{equation}
\begin{split}
\hat{H}_\Diamond&=\sum_{i,j=1}^34E_C^{(ij)}\left(\hat{n}_i-n_g^{(i)}\right)\left(\hat{n}_j-n_g^{(j)}\right) + \\
&-\sum_{i=1}^3E_J^{(i)}\cos\left(\hat{\varphi}_i\right)-E_J^{(4)}\cos\left(\sum_{i=1}^3\hat{\varphi}_i - \frac{2\pi\Phi_\mathrm{ext}}{\Phi_0}\right),
\end{split}
\label{eq:rhombi_hamiltonian}
\end{equation}
where $E_C^{(ij)}$ are the self- and cross-charging energies associated with the three modes, $\hat{n}_i$ and $e^{\mathrm{i}\hat\varphi_i}$ are the conjugate Cooper pair number and (periodically defined) phase operators, with $[\hat n_i,e^{\mathrm{i}\hat\varphi_j}]=e^{\mathrm{i}\hat\varphi_i}\delta_{ij}$, $n_g^{(i)}$ are the offset charges, and $E_J^{(i)}$ are the Josephson energies. All phase coordinates are periodic on $[-\pi,\pi)$, defining a cubic configuration space. At frustration, the potential energy features a three-dimensional double-well potential (Fig.~\ref{fig:fig3}a), where, for equal junction energies, the two minima are located symmetrically around the origin at the coordinates of $\varphi_1^0=\varphi_2^0=\varphi_3^0=\pm\pi/4$. The locations of these minima are essential for the charge-parity protection. For example, if the junctions are unequal, the potential still has a double-well structure, but the minima are slightly shifted. 

To shed light on the structure of the wavefunctions, it is beneficial to examine a nonphysical case first, when we approximate the phase wavefunctions of the computational basis states as symmetric and antisymmetric superpositions of two Dirac deltas located at the two minima of the potential. In this case, the wavefunctions in charge space (ignoring normalization) are
\begin{equation}
    \begin{split}
    \tilde\psi_0(n_1,n_2,n_3) &= \cos\left(\frac{n_1 + n_2 + n_3}{4}\pi\right), \\
    \tilde\psi_1(n_1,n_2,n_3) &= \sin\left(\frac{n_1 + n_2 + n_3}{4}\pi\right).
    \end{split}
\end{equation}
These states, also shown in Fig.~\ref{fig:fig3}a, display that certain combinations of charge states carry no spectral weight, a signature of an Aharonov-Bohm-type interference. We note that when the minima are slightly away from the $\pm\pi/4$ values (due to junction asymmetries), these prohibited charge states are not present.

Next, employing numerical calculations, we obtain the wavefunctions for more realistic cases. First, when equal capacitors shunt each junction (Fig.~\ref {fig:fig3}b), the ground and excited states display a similar interference pattern to the unphysical states in Fig.~\ref{fig:fig3}a, with certain prohibited charge states. Second, when a large capacitor is placed across two opposite nodes, the charge-parity structure is even more apparent. In this case, even/odd charge states along the $n_1=n_2$ axis can be observed (Fig.~\ref{fig:fig3}c). From a different view, the wavefunctions in phase space become squeezed in one direction (due to the strongly anisotropic capacitance values) and an approximate $\pi$-phase shift symmetry is recovered.  These observations demonstrate how a multi-mode rhombus approximates a single-mode $\cos(2\varphi)$ circuit, when a large shunt capacitor is used across the device.

\begin{figure}[t]
\centering
\includegraphics[width = 8.6cm]{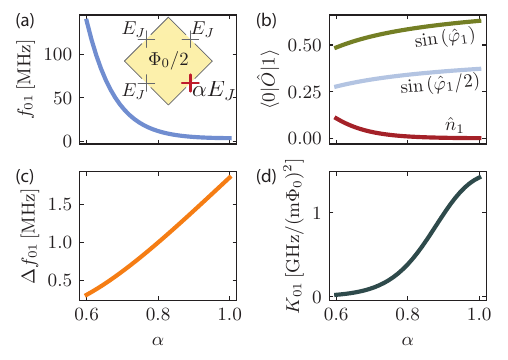}
\caption{\label{fig:fig4} \textbf{Numerical simulation of the properties of the soft-rhombus qubit} (a) The qubit frequency at frustration as a function of the junction asymmetry $\alpha$. The asymmetry increases the transition frequency, leading to lower $1/f$ flux noise amplitude at $f_{01}$. (b) Various matrix elements of the computational states as a function of $\alpha$. While asymmetry increases the charge matrix element, other matrix elements reduce. (c) The charge dispersion of the qubit transition, which is calculated as the change in frequency when $n_g^{(1)}$ is changed by half a period. The qubit becomes less charge-noise sensitive as the asymmetry is increased. (d) The curvature of the qubit frequency vs.~flux at frustration plotted as a function of asymmetry. The lower curvature leads to lower susceptibility to second-order flux noise dephasing. In the simulation, we used the circuit parameters of the real device.}
\end{figure}

\subsection{The soft-rhombus qubit}

A key aspect of the rhombus is that the emergence of charge-parity symmetry does not guarantee protection against all relevant relaxation mechanisms. In particular, the $\sin(\hat{\varphi}_i)$ or $\sin(\hat{\varphi}_i/2)$ operators can couple the computational states, leading to loss in the presence of flux noise or quasiparticle loss. The loss rates are further amplified by the low transition frequency of the qubit, where, for example, the spectral density of $1/f$-flux noise has the highest amplitude. A possible route to address this issue is to raise the frequency of the qubit and partially sacrifice the protection against processes associated with the $\hat{n}_i$ operators. We achieve this by lowering the energy of one of the junctions compared to the other junctions with a ratio of $\alpha$, such that $E_J^{(4)}<E_J^{(1)}\approx E_J^{(2)}\approx E_J^{(3)}$, which moves the two potential minima closer in the three-dimensional phase space. As the numerical simulations in Fig.~\ref{fig:fig4} shows, this approach has several benefits: (1) the qubit has a higher frequency and experiences less of a $1/f$ flux noise amplitude (Fig.~\ref{fig:fig4}a), (2) the magnitude of the matrix elements of the $\sin(\hat{\varphi}_i)$ and $\sin(\hat{\varphi}_i/2)$ operators get slightly reduced (Fig.~\ref{fig:fig4}b), (3) the charge sensitivity of the device becomes more suppressed (Fig.~\ref{fig:fig4}c), and (4) the curvature of the transition with respect of flux is lowered (Fig.~\ref{fig:fig4}d). These observations suggest that the modified version of the rhombus, which we have nicknamed the soft-rhombus qubit, can potentially outperform the ideal case of the rhombus, as long as the noise associated with the $\hat{n}_i$ operators is not the dominant source of loss at low frequencies.

\subsection{Experimental realization of the soft-rhombus}

\begin{figure}[t]
\centering
\includegraphics[width = 8.6cm]{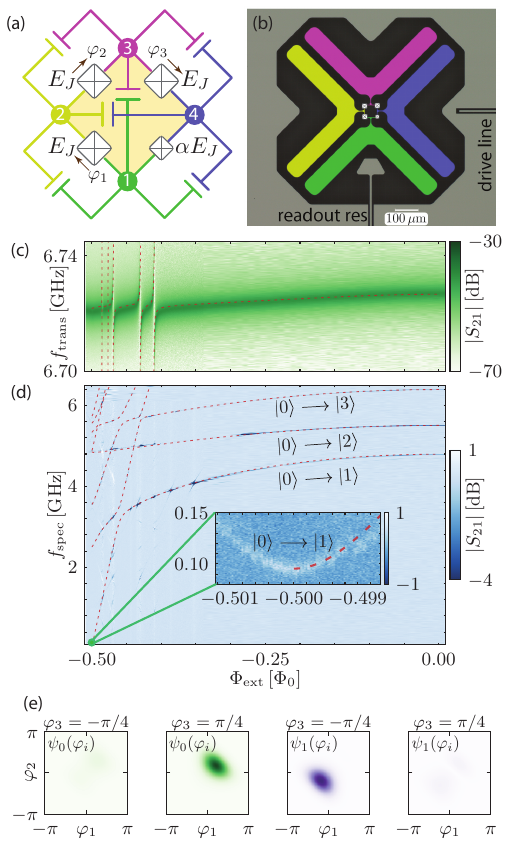}
\caption{\label{fig:fig5} \textbf{The experimentally realized soft-rhombus qubit} (a) Circuit schematics of the soft-rhombus qubit, where one of the junctions has a different energy ($\alpha E_J$) than the other ones ($E_J$). (b) False color image of the soft-rhombus qubit that is capacitively coupled to a readout resonator and a drive line. Josephson junction symbols indicate the position of the physical junctions. (c) The transmission of the readout resonator as a function of external flux, when a single tone $f_\mathrm{trans}$ is swept. The fit is plotted on top of the data as red dashed lines. (d) The measured energy spectrum of the qubit as a function of external flux, which is detected as a change in the transmission of the readout resonator when a spectrocopy tone $f_\mathrm{spec}$ is swept.  The inset shows an enlarged region of the $|0\rangle\longrightarrow|1\rangle$ transition around frustration. (e) Slices of the three-dimensional phase wavefunctions of the ground and excited state at $\Phi_\mathrm{ext}=0.49\Phi_0$, showing that these states have disjoint supports.}
\end{figure}

Motivated by these findings, we experimentally realized a soft-rhombus qubit, where one of the junctions is intentionally reduced by a ratio of $\alpha=0.63$ (Fig.~\ref{fig:fig5}a). The device was fabricated on a sapphire substrate with tantalum capacitor pads and traditional double-angle evaporated aluminum-oxide-based Josephson junctions (Methods~\ref{app:fabrication}). The qubit is capacitively coupled to a $\lambda/2$ resonator for read-out, and to a transmission line for coherent control (Fig.~\ref{fig:fig5}b). The symmetric four large capacitor pads ensure that the qubit transitions are offset-charge insensitive. The details of the device parameters are discussed in Methods~\ref{app:cq}. The qubit is flux-biased by an external magnetic coil and measured at the base temperature of a dilution refrigerator.

To obtain the energy-flux dispersion of the qubit, we carry out standard one- and two-tone resonator spectroscopy. First, we measure the resonator at low photon numbers as a function of flux, which displays avoided crossings when transitions of the rhombus are in resonance with the resonator (Fig.~\ref{fig:fig5}c). Then, we monitor the response of the resonator as a second (spectroscopic) tone is swept. When the spectroscopic tone is on-resonance with a rhombus transition, we detect a change in the transmission of the read-out resonator due to dispersive interaction. Figures \ref{fig:fig5}c and d also show the fit of the spectroscopy data using a standard coupled resonator-qubit model (Methods~\ref{app:cq}.) We find an excellent agreement between the predicted transitions of the multi-mode circuit and the experimental data. While around zero external flux, we see three plasmon-like transitions (corresponding to the transitions of the trimon qubit~\cite{PhysRevApplied.7.054025, maurya2026}), closer to half flux quantum, we detect the bit-flip-protected fluxon states of the rhombus. The protection of these states arises because the qubit wavefunctions are localized in different valleys in the three-dimensional phase space and have disjoint support (Fig.~\ref{fig:fig5}e). Exactly at half flux quantum, we observe a first-order flux-insensitive sweet spot (slightly below 100\,MHz), where the qubit eigenstates are a symmetric and an antisymmetric combination of states localized in the two valleys.

\subsection{Noise analysis of the rhombus}

\begin{figure}[t]
\centering
\includegraphics[width = 8.6cm]{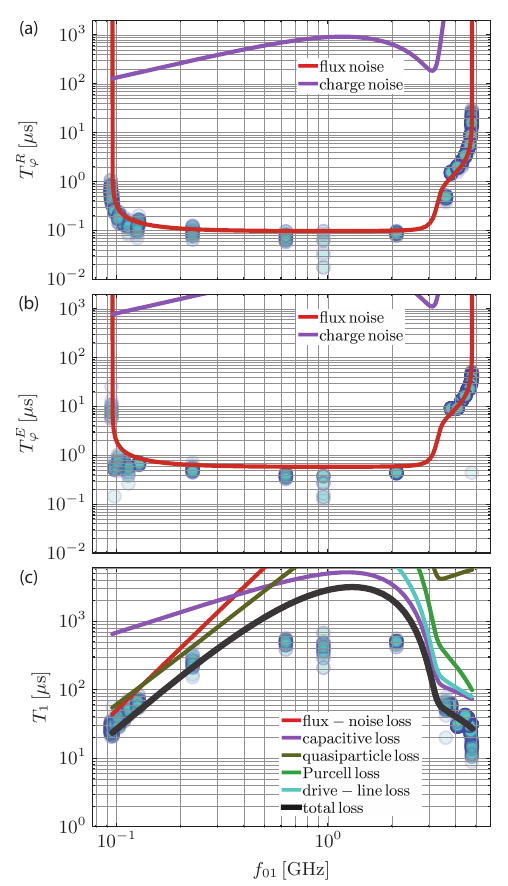}
\caption{\label{fig:fig6} \textbf{Measured and predicted dephasing and relaxation times} (a), (b) The Ramsey and echo pure dephasing times measured at different qubit frequencies $f_{01}$ in the soft-rhombus qubit. The different markers at a given flux value show different measurements obtained over several hours. The solid lines display theoretical expectations based on a $1/f$ charge and flux noise model. The flux noise amplitude is $A_\Phi=4\,\mu\Phi_0$, while the charge noise amplitude is $A_n=2\times10^{-4}$. The ratio between the Ramsey and echo dephasing rates is $\chi=6$. (c) The measured relaxation times as a function of frequency (markers). Solid lines show the expectation of different noise models. The flux noise amplitude is $A_\Phi=4\,\mu\Phi_0$, the capacitive quality factor is $Q_C=8\times10^{5}$, the quasiparticle density is $x_\mathrm{qp}=10^{-8}$, and the temperature is $T=50\,$mK.}
\end{figure}

We characterize the relaxation and dephasing times of the soft-rhombus qubit using calibrated $\pi$ and $\pi/2$ pulses. For the relaxation time $T_1$ measurements, we fit the time-dependence of the resonator response by an exponential decay $e^{-t/T_1}$, while for the pure dephasing times $T_\varphi$, we fit the measured curves with a Gaussian decay with the function of $e^{-t/2T_1}e^{-(t/T_\varphi)^2}\cos(2\pi f t + \phi)$, where $f$ is the frequency of the change of the phase of the second $\pi/2$ pulse. Regarding the Ramsey dephasing times (Fig.~\ref{fig:fig6}a), we obtain the largest values at the first-order flux-insensitive points, such as around zero flux (a few tens of microseconds) and around frustration (around half a microsecond). Away from these points, the dephasing times plummet below 100\,ns. The echo measurements show similar behavior, with the best pure dephasing times in the tens of microseconds (Fig.~\ref{fig:fig6}b). Regarding the relaxation times, we measure $T_1$ values in the range of a few tens of microseconds for the plasmon states, whereas a few hundreds of microseconds for the fluxon states around a GHz qubit frequencies (Fig.~\ref{fig:fig6}c). As the frequency is decreased, however, we observe a significant reduction in the relaxation times to the tens of microseconds regime. 

To understand the noise sources that limit the performance of the qubit, we extend noise models based on single-mode circuits~\cite{Schoelkopf2003,smith_superconducting_2020}. Our treatment is detailed in Methods~\ref{app:noise_models}, and here we summarize the results. Due to its multi-mode structure, the circuit couples to multiple noise sources, including three fluctuating gate voltages, an oscillating external flux, and a lossy quasiparticle bath. Regarding the pure dephasing times, the Ramsey and echo data are in excellent agreement with the theoretical prediction of a standard flux noise model with flux noise amplitude of $A_\Phi=4\,\mu\Phi_0$ over the entire range of the qubit transitions (red lines in Fig.~\ref{fig:fig6}a,b). The sensitivity of the device to charge noise is negligible due to the four large capacitor pads (purple lines in Fig.~\ref{fig:fig6}a,b). Thus, given the excellent agreement with the flux-noise model, these measurements provide us with a precise value for the flux noise amplitude in our circuit that we can feed into the relaxation models. 

Regarding the relaxation times, we use Fermi's golden rule to obtain the losses due to various noise channels (Fig.~\ref{fig:fig6}c). At higher frequencies, we find that the combination of capacitive loss and the coupling to the $50\,\Omega$ environment through the drive line and the resonators limits the performance of the plasmon states. In contrast, at low frequencies, we find that flux noise with the noise amplitude extracted from the pure dephasing times and quasiparticle loss dominate the relaxation processes. This further highlights the advantage of the higher operation frequency in the soft-rhombus regime over the symmetric rhombus. We note that while the theoretical loss model captures the trend in the measured relaxation times, there are discrepancies present. On one hand, at lower frequencies, the device performs slightly better than expected, which suggests that the used quasiparticle model overestimates the relaxation rates, possibly due to correlations between quasiparticle tunneling events. On the other hand, at intermediate frequencies, our noise models underestimate the observed relaxation rates, implying that other loss mechanisms, such as heating to higher levels closer to the resonator, can play a role~\cite{PhysRevX.11.011010}.

\section{Discussion}

The rhombus circuit is one of the archetypal protected qubits, which can be harnessed for topologically protected array qubits, when multiple of these circuits are concatenated and biased at half a flux quantum. In this work, we analyzed a single rhombus using a full circuit description and noise model to understand the degree of protection and identify an optimal parameter regime for this multi-mode qubit. Surprisingly, we find that moving away from the charge-parity protected regime is beneficial because coupling to flux and quasiparticle noise can be reduced by introducing asymmetries in the junctions. This stems from the fact that the rhombus qubit is only protected against relaxation processes associated with local operators in charge space, while flux and quasiparticle noise can couple states with different charge parity. Furthermore, we identify a biased-noise regime of the circuit away from flux frustration, where the qubit wavefunctions are localized around different minima in phase space, leading to long relaxation times at the expense of short dephasing times. 

We have experimentally realized this descendant of the charge-parity protected rhombus qubit, where one of the junctions has intentionally different energy than the other ones. In this soft-rhombus qubit, the quasi-degeneracy of the qubit states at frustration is lifted, which allows us to address the low-frequency qubit transition directly. By studying the energy dependence of the dephasing and relaxation rates, we verified that flux noise and quasiparticle-tunneling-induced processes are the main limiting sources for relaxation at low frequencies. Both of these losses are reduced away from frustration, where the soft-rhombus qubit can be operated as a biased-noise qubit with relaxation times in the regime of hundreds of microseconds. These findings highlight that low-frequency noise is a critical issue for certain protected qubits, and taking compromises based on realistic noise sources is a key consideration for their design.

While preparing this manuscript, we became aware of a similar work that investigates the tunable version of the rhombus qubit and arrives at similar conclusions regarding the loss mechanisms affecting interferometer-based circuits~\cite{delft_paper}.

\section*{Acknowledgments}
We thank Pranav Mundada for the inspiring initial conversations on the rhombus qubit. We also thank Zhenxing Liu for helpful discussions in the early stages of this project. This work was supported by the University of Colorado Boulder Research \& Innovation Seed Grant Program and the NSF Faculty Early Career Development (CAREER) Program under Award Numbers 2440002 and 2240129. Certain commercial instruments are identified to specify the experimental study adequately. This does not imply endorsement by NIST or that the instruments are the best available for the purpose. T.F.Q.L acknowledges support in part by an appointment to the NRC Research Associateship Program at the National Institute of Standards and Technology, administered by the Fellowships Office of the National Academies of Sciences, Engineering, and Medicine.

\section*{Methods}
\appendix

\section{Device Fabrication}
\label{app:fabrication}
\begin{figure}[h]
    \centering
    \includegraphics[width=8.6cm]{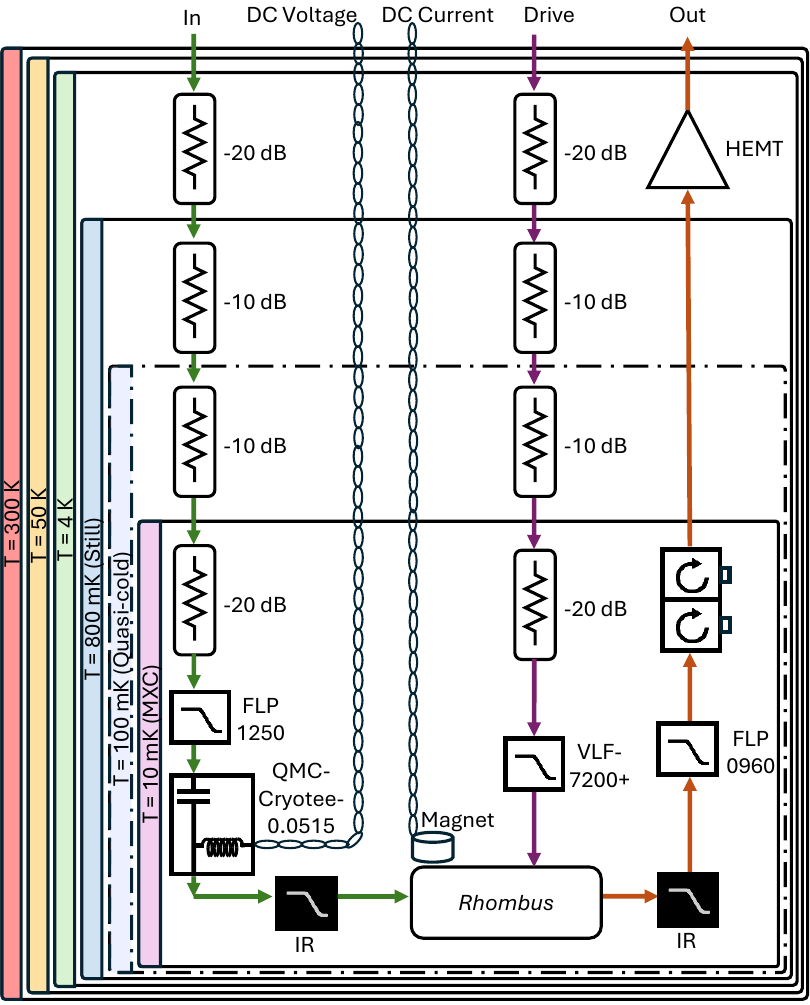}
    \caption{Wiring diagram of the measurement setup inside the dilution refrigerator.}
    \label{fig:wiring}
\end{figure}

The device was fabricated on a c-plane sapphire substrate. The large features, such as the capacitor pads, the ground plane, the readout resonator, and the drive line were fabricated from epitaxial 195\,nm thick tantalum films grown on 5\,nm thick niobium nucleation layer ~\cite{Tony.tantalum}. The base layer was patterned with photoresist (Microposit MF-26A and HDMS for adhesion promotion) and etched with an SF$_{6}$ process. The resist was removed with an NMP-based solvent, then the wafer was rinsed in isopropanol and exposed to a 2-minute 6:1 buffered oxide etch. A stack of 380\,nm thick PMGI and 125\,nm thick PMMA was used for the e-beam lithography step (with an additional Electra 92 e-beam resist layer to prevent charging). After e-beam exposure, the Electra layer was removed with DI water, and the PMMA layer was developed in a standard cold 3:1 IPA/DI water mixture, followed by developing the undercut layer in room-temperature PMGI 101A. A brief 20-second oxygen plasma was used to clean the substrate before junction deposition, as well as a 1-minute in-situ ion milling. The double-angle junction deposition consisted of an initial 45\,nm aluminum film deposition at 45$^{\circ}$ tilt, a 10-minute oxidation at 565\,mTorr, and a 95\,nm film deposition at 0$^{\circ}$ tilt. The top of the aluminum layer was subsequently exposed to a 10-minute 5\,Torr oxidation for clean surface termination. After deposition, the wafer was diced, and individual chips were lifted off in 70$^{\circ}$C PG remover and rinsed in IPA. Following liftoff, the chips were further cleaned with a 15-minute ozone cleaning procedure to remove the e-beam resist residue around the junction area. Finally, the chips were wirebonded, packaged, and cooled down in a dilution refrigerator (Fig.~\ref{fig:wiring}).

\section{Circuit quantization on the rhombus qubit}
\label{app:cq}

\begin{figure}
    \centering
    \includegraphics{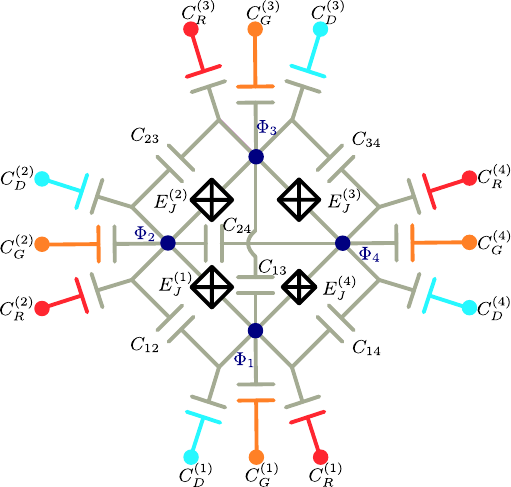}
    \caption{Full capacitances of the Rhombi, including capacitance to ground, readout resonator, and drive. Also shown are the node fluxes and junctions.}
    \label{fig:capacitance}
\end{figure}

Here, we provide details on the derivation of the Hamiltonian of the rhombus. Figure \ref{fig:capacitance} shows the circuit diagram, including the coupling capacitors. The capacitance between nodes $i$ and $j$ are $C_{ij}$, while $C_G^{(i)}$, $C_R^{(i)}$ and $C_D^{(i)}$ are the capacitance values between node $i$ and the ground, the centerpin of the readout resonator, and the drive line, respectively, First, we introduce the vector of the four generalized node flux variables of the rhombus qubit as $\mathbf{\Phi}=\left[\Phi_1, \Phi_2, \Phi_3, \Phi_4\right]^T$. The circuit couples to three different external voltage sources: to the centerpin of the resonator with voltage $V_R$, to the centerpin of the drive line with voltage $V_D$, and to the ground plane, which defines the zero potential $V_G=0$. Accordingly, we introduce the voltage vectors $\mathbf{V}_R=[V_R, V_R, V_R, V_R]^T$, and $\mathbf{V}_D=[V_D, V_D, V_D, V_D]^T$. The capacitance matrix of the rhombus is
\begin{equation*}
\mathbf{C}^\Phi=
    \begin{pmatrix}
    +C_{11} & -C_{12} & -C_{13} & -C_{14}\\
    -C_{12} & +C_{22} & -C_{23} & -C_{24}\\
    -C_{13} & -C_{23} & +C_{33} & -C_{34}\\
    -C_{14} & -C_{24} & -C_{34} & +C_{44}
    \end{pmatrix},
\end{equation*}
where $C_{ii}=C_G^{(i)}+C_R^{(i)}+C_D^{(i)}+\sum_{j\ne i=1}^{4} C_{ij}$. When we write the resonator and drive coupling capacitance matrix as $\mathbf{C}_R^\Phi=\text{diag}[C_R^{(1)}, C_R^{(2)}, C_R^{(3)}, C_R^{(4)}]$ and $\mathbf{C}_D^\Phi=\text{diag}[C_D^{(1)}, C_D^{(2)}, C_D^{(3)}, C_D^{(4)}]$, the Lagrangian of the rhombus qubit reads
\begin{equation*}
    L_\Phi=\frac{1}{2}\Dot{\mathbf{\Phi}}^T\mathbf{C}^\Phi\Dot{\mathbf{\Phi}}-\Dot{\mathbf{\Phi}}^T\mathbf{C}_R^\Phi\mathbf{V}_R-\Dot{\mathbf{\Phi}}^T\mathbf{C}_D^\Phi\mathbf{V}_D-U(\mathbf{\Phi}, \Phi_\mathrm{ext}),
\end{equation*}
where $U(\mathbf{\Phi}, \Phi_\mathrm{ext})$ is the potential energy of the four junctions.

Next, we move to the basis of branch fluxes, using the transformation
\begin{equation*}
    \mathbf{\Theta}=\mathbf{T}\cdot\mathbf{\Phi},
\end{equation*}
where the transformation matrix is
\begin{equation*}
    \mathbf{T}=
    \begin{pmatrix}
    -1 & 1 & 0 & 0\\
    0 & -1 & 1 & 0\\
    0 & 0 & -1 & 1\\
    \varsigma_1 & \varsigma_2 & \varsigma_3 & \varsigma_4
    \end{pmatrix}
\end{equation*}
and $\varsigma_i=(C_R^{(i)}+C_D^{(i)}+C_G^{(i)})/(\sum_j(C_R^{(j)}+C_D^{(j)}+C_G^{(j)}))$.

After the transformation, the Lagrangian reads
\begin{equation*}
    L_\Theta=\frac{1}{2}\Dot{\mathbf{\Theta}}^T\mathbf{C}^\Theta\Dot{\mathbf{\Theta}}-\Dot{\mathbf{\Theta}}^T\mathbf{C}_R^\Theta\mathbf{V}_R^\Theta-\Dot{\mathbf{\Theta}}^T\mathbf{C}_D^\Theta\mathbf{V}_D^\Theta-U(\mathbf{T}^{-1}\mathbf{\Theta}, \Phi_\mathrm{ext}),
\end{equation*}
where $\mathbf{C}^\Theta = [\mathbf{T}^{-1}]^T\mathbf{C}^\Phi\mathbf{T}^{-1}$, $\mathbf{C}_R^\Theta = [\mathbf{T}^{-1}]^T\mathbf{C}_R^\Phi\mathbf{T}^{-1}$, $\mathbf{C}_D^\Theta = [\mathbf{T}^{-1}]^T\mathbf{C}_D^\Phi\mathbf{T}^{-1}$, $\mathbf{V}_R^\Theta=\mathbf{T}\mathbf{V}_R$, and $\mathbf{V}_D^\Theta=\mathbf{T}\mathbf{V}_D$.

Next, we perform a Legendre transformation, where we define the charge variables as $\mathbf{q}_\Theta={\partial L_\Theta}/{\partial \Dot{\mathbf{\Theta}}}$, and obtain the Hamiltonian of the rhombus qubit 
\begin{equation}
\begin{split}
    {H}_\Diamond&=\\
    &\frac{1}{2}[\mathbf{q}_\Theta+\mathbf{C}_R^\Theta \mathbf{V}_R^\Theta +\mathbf{C}^\Theta_D\mathbf{V}^\Theta_D]^T[\mathbf{C}^\Theta]^{-1} \times\\
    &\times[\mathbf{q}_\Theta+\mathbf{C}_R^\Theta \mathbf{V}_R^\Theta +\mathbf{C}^\Theta_D\mathbf{V}^\Theta_D]+U(\mathbf{T}^{-1}\mathbf{\Theta}, \Phi_\mathrm{ext}).
\end{split}
\end{equation}

The coupling of the circuit to the resonator and the drive line can be further expressed with the $\beta$ coefficients as 
\begin{equation*} 
\begin{split}
H_{\text{R}}&=\mathbf{q}_\Theta^T[\mathbf{C}^\Theta]^{-1}\mathbf{C}^\Theta_R\mathbf{V}^\Theta_R=\sum_{i=1}^3\beta_R^{(i)}q_iV_R, \\
H_{\text{D}}&=\mathbf{q}_\Theta^T[\mathbf{C}^\Theta]^{-1}\mathbf{C}^\Theta_D\mathbf{V}^\Theta_D=\sum_{i=1}^3\beta_D^{(i)}q_iV_D.
\end{split}
\end{equation*}

When moving into the quantum description of the circuit, we introduce the offset charges and promote the classical variables to quantum operators to arrive at Eq.~\ref{eq:rhombi_hamiltonian}. The data was fit by a constrained optimization fitting using a coupled circuit-resonator model 
\begin{equation}
    \hat{H}_\mathrm{total} = \hat{H}_\Diamond  + \hbar\omega_\mathrm{res}a^\dagger a + 2eV_0\sum_{i=1}^3 \beta^{(i)}_R\hat{n}_i\left(\hat{a} + \hat{a}^\dagger\right),
\end{equation}
where $a$ and $a^\dagger$ are the ladder operators of the photons in the readout resonator with frequency $\omega_\mathrm{res}/2\pi$, $V_0= \sqrt{\hbar \omega_\mathrm{res}^2Z_R/\pi}$ is the zero-point fluctuation for the $\lambda/2$ resonator, $Z_R=50\,\Omega$ is the nominal impedance of the resonator, and $\beta^{(i)}_R$ describes the coupling coefficient between mode $i$ and the resonator. The fitted parameters of circuit are $E_C^{(11)}/h = E_C^{(22)}/h = E_C^{(33)}/h = 275.8\,$MHz, $E_C^{(12)}/h =E_C^{(21)}/h=E_C^{(23)}/h =E_C^{(32)}/h = 115.4\,$MHz, $E_C^{(13)}/h =E_C^{(31)}/h=46.5\,$MHz, $E_J^{(1)}/h=13.04\,$GHz, $E_J^{(2)}/h=13.12\,$GHz, $E_J^{(3)}/h=12.92\,$GHz, $E_J^{(4)}/h=8.20\,$GHz, $\beta_R^{(1)}=-0.0813$, $\beta_R^{(2)}=-0.0197$, $\beta_R^{(3)}=0.0193$, and the resonator frequency is $f_\mathrm{res}=6.7198\,$GHz.

\section{Loss models of the soft-rhombus}
\label{app:noise_models}

In this section, we describe the loss models used to predict the relaxation and dephasing rates in the soft-rhombus qubit. In general, the coupling Hamiltonian between a classical noise source and the qubit can be written as $\hat{H}_c=\hat{\mathcal{O}}\cdot\xi(t)$, where $\hat{\mathcal{O}}$ is a qubit operator, and $\xi(t)$ is a noisy external parameter. Such coupling can cause transitions between the qubit states (relaxation) and changes in the qubit frequency (dephasing).

To estimate the relaxation times, we follow Fermi's golden rule, which states that the relaxation rate associated with $\xi$ is
\begin{equation}
    \frac{1}{T_1^\xi} = \Gamma_1^\xi = \frac{1}{\hbar^2}\left|\langle 0|\hat{\mathcal{O}}|1\rangle\right|^2\cdot S_{\xi}^+(\omega),
\end{equation}
when both emission and absorption processes are included. Here, $|0\rangle$ and $|1\rangle$ are the qubit states, $S_{\xi}^+(\omega)=S_{\xi}(\omega) + S_{\xi}(-\omega)$ is the symmetric noise spectral density of the bath, and $S_{\xi}(\omega)=\int_{-\infty}^{+\infty}dt \langle\xi(0)\xi(t)\rangle e^{-i\omega t}$. 

Regarding pure dephasing, we calculate the average of the random phase accumulation that results from the changing qubit frequency. First, we divide the external parameter into a time-independent static bias value $\xi_0$ and an oscillating term $\delta\xi(t)$, such that $\xi(t) = \xi_0 + \delta\xi(t)$. This leads to a change in the qubit frequency 
\begin{equation}
    \omega_{01}[\xi(t)]\approx \omega_{01}(\xi_0) + \frac{\partial\omega_{01}}{\partial\xi}\cdot\delta\xi(t).
\end{equation}

In the case of a $1/f$-type of noise, the noise spectral density can be written as $S_\xi(\omega)=2\pi A_\xi^2/|\omega|$, where $A_\xi$ is the amplitude of the noise. The dephasing rates obtained through Ramsey and echo measurements are proportional to the noise amplitude and the sensitivity of the qubit to the external parameter~\cite{PhysRevB.72.134519}
\begin{equation}
    \begin{split}
        \Gamma_{R,\varphi}^\xi &= A_\xi \cdot  \left|\frac{\partial\omega_{01}}{\partial\xi}\right|\cdot\sqrt{O(1)-\ln(\omega_\mathrm{IR}t)} \\
       \ \Gamma_{E,\varphi}^\xi &= A_\xi \cdot  \left|\frac{\partial\omega_{01}}{\partial\xi}\right|\cdot\sqrt{\ln2},
    \end{split}
\end{equation}
where the factor of $\sqrt{O(1)-\ln(\omega_\mathrm{IR}t)}$ depends on time $t$ and the infrared cutoff $\omega_\mathrm{IR}$, which are determined by the measurement protocol. Here, we assume that the measured dephasing times are connected through a constant $\chi>1$, such that $\Gamma_{R,\varphi}^\xi=\chi\cdot\Gamma_{E,\varphi}^\xi$, as commonly done in the field~\cite{Bylander2011, PRXQuantum.5.030341}.

\subsubsection{Dephasing due to charge and to flux noise}

As the Hamiltonian in Eq.~\ref{eq:rhombi_hamiltonian} shows, the rhombi qubit couples to three offset charges $n_{g}^{(i)}$ ($i=1,2,3$). We approximate the charge-sensitivity of the device with the magnitude of the following gradient 
\begin{equation}
    \left|\nabla_{n_{g}}\omega_{01}\right|\approx \sqrt{\sum_{i=1}^3\left(\frac{\partial\omega_{01}}{\partial n_g^{(i)}}\right)^2 },
\end{equation}
where
\begin{equation}
    \frac{\partial\omega_{01}}{\partial n_{g}^{(i)}}\approx \frac{\omega_{01}(n_{g}^{(i)}=0.5) - \omega_{01}(n_{g}^{(i)}=0.0)}{0.5}.
\end{equation}

Taking a typical value for the charge noise amplitude $A_n=2\times10^{-4}$~\cite{Charge.noise.amplitude}, the predicted dephasing rates are orders of magnitude larger than the measured values, hence the device is not limited by charge noise (Fig.~\ref{fig:fig6}a,b).

The rhombi qubit has a single flux loop, which couples to an external flux $\Phi_\mathrm{ext}$. We obtain the sensitivity of the device to flux noise by numerically calculating $|\partial\omega_{01}/\partial\Phi_\mathrm{ext}|$. Using a typical flux noise amplitude of $A_\Phi=4\,\mu\Phi_0$ and a ratio of $\chi=6$, we find a good agreement between the measured Ramsey and echo dephasing rates over a large range of energies (Fig.~\ref{fig:fig6}a,b), indicating that the limiting factor for the pure dephasing in the rhombus is flux noise.

\subsubsection{Relaxation due to dielectric loss}

Dielectric or capacitive loss arises from the fluctuating voltages in the materials surrounding the superconducting capacitor pads. Here, we model this by noisy gate voltages affecting the qubit. By expressing the gate voltages as $n_g^{(i)}(t) = n_{g,0}^{(i)} + \delta n_g^{(i)}(t)$ ($i=1,2,3$), the coupling Hamiltonian takes the form
\begin{equation}
    \hat{H}_c^\mathrm{cap}=-8\sum_{i=1}^3 E_{C}^{(ii)}\hat{N}_C^{(i)}\delta n_g^{(i)}(t),
\end{equation}
where we introduced the combined charge operators
\begin{equation}
    \hat{N}_C^{(i)} =  \sum_{j=1}^3\frac{E_{C}^{(ij)} + E_{C}^{(ji)}}{2E_{C}^{(ii)}}\hat{n}_j.
\end{equation}

We obtain the relaxation rate due to this coupling by generalizing the results on the loss rate of a single mode. When a mode with charge operator $\hat{n}$ containing a single capacitance $C$ with charging energy $E_C=e^2/2C$ and quality factor $Q_C$ is coupled to a noisy voltage source $V(t)$, the coupling Hamiltonian is $H_c=2e\hat{n}\cdot V(t)=8E_C\hat{n}\cdot n_g(t)$, where $n_g(t)=CV(t)/2e$. The noise spectral density is~\cite{smith_superconducting_2020}
\begin{equation}
    S_V^+(\omega)=\frac{2\hbar}{Q_CC}\coth\left(\frac{\hbar|\omega|}{2k_BT}\right)=\frac{4E_C\hbar}{e^2Q_C}\coth\left(\frac{\hbar|\omega|}{2k_BT}\right).
\end{equation}
Thus, in the case of the soft-rhombus qubit, the relaxation rate due to the dielectric loss can be approximated as
\begin{equation}
    \Gamma_1^\mathrm{diel} = \frac{32\pi}{Q_C}\sum_{i=1}^3\left[\frac{E_C^{(ii)}}{h}\left|\langle 0|\hat{N}_C^{(i)}|1\rangle\right|^2\right]\coth\left(\frac{\hbar\omega_{01}}{2k_BT}\right).
\end{equation}

As the purple colored lines in Fig.~\ref{fig:fig6}c show, this type of noise mostly affects the transmon transitions in the soft-rhombus around zero flux, while it is not a limiting factor for the fluxon states close to frustration. 

\subsubsection{Drive-line relaxation and Purcell effect}

The soft-rhombus qubit couples via all its three modes to the noisy $Z_0=50\,\Omega$ environment, both directly through the drive line and indirectly through the resonator. The coupling to the drive line with voltage bias $V_D(t)$ is described by the $\beta_D^{(i)}$ coefficients, which we obtain from finite-element capacitance matrix simulations. The coupling Hamiltonian is
\begin{equation} \hat{H}_c^\mathrm{drive}= 2e \hat{N}_DV_D(t),
\end{equation}
where
\begin{equation}
    \hat{N}_D = \sum_{i=1}^3 \beta_{D}^{(i)}\hat{n}_i.
\end{equation}
The noise spectral density of the 50\,$\Omega$ environment can be expressed as
\begin{equation}
    S_V^+(\omega)=2Z_0\hbar\omega\coth\left(\frac{\hbar\omega}{2k_BT}\right).
\end{equation}
This leads to the relaxation rate of
\begin{equation}
    \Gamma_1^\mathrm{drive} = \frac{16\pi Z_0e^2}{h}\left|\langle 0|\hat{N}_D|1\rangle\right|^2 \omega_{01}\coth\left(\frac{\hbar\omega_{01}}{2k_BT}\right).
\end{equation}

In addition, the relaxation through the resonator can be approximated for the plasmon-like states by the linewidth $\kappa$ of the resonator (obtained from RF simulation), the coupling strength between the qubit and the resonator $g$ (subtracted from the fit to the spectroscopy data), and the detuning between the qubit and the resonator~\cite{PhysRevLett.101.080502}
\begin{equation}
    \Gamma_1^\mathrm{Purcell} = \kappa\left(\frac{g}{\omega_\mathrm{res}-\omega_{01}}\right)^2.
\end{equation}

The estimated loss due to these two effects mostly influences the performance of plasmon states at higher frequencies (green and blue lines on Fig.~\ref{fig:fig6}c).

\subsubsection{Relaxation due to flux noise}

The rhombi qubit has a flux loop, hence it is subject to flux noise. A fluctuating magnetic field can arise from several sources, such as external magnetic noise or uncompensated spins on the surface of the superconductor~\cite{PhysRevApplied.13.054079}.  Assuming a small fluctuating magnetic flux $\delta\Phi_\mathrm{ext}(t)$ around a constant flux value $\Phi_\mathrm{ext}^{0}$, and a symmetric allocation of the external flux, the approximate coupling Hamiltonian is 
\begin{equation}
    \hat{H}_c^\mathrm{flux} = \hat{\mathcal{O}}_\Phi\cdot \delta\Phi_\mathrm{ext}(t),
\end{equation}
where 
\begin{equation}
\begin{split}
    & \hat{\mathcal{O}}_\Phi = \frac{\pi}{2\Phi_0} \times\\ & \left[\sum_{i=1}^3E_J^{(i)}\sin\left(\hat\varphi_i+\frac{\pi\Phi_\mathrm{ext}^{0}}{2\Phi_0}\right)-  E_J^{(4)}\sin\left(\sum_{i=1}^3\hat\varphi_i  -\frac{\pi\Phi_\mathrm{ext}^{0}}{2\Phi_0}\right)\right].
\end{split}
\end{equation}

Using the expression for the noise spectral density for $1/f$ noise $S_{\Phi_\mathrm{ext}}(\omega)=2\pi A_\Phi^2/|\omega|$, the relaxation time due to flux noise in the rhombus is
\begin{equation}
    \Gamma_1^\mathrm{flux} = \frac{1}{\hbar^2}\left|\langle 0|\hat{\mathcal{O}}_\Phi|1\rangle\right|^2\frac{4\pi A_\Phi^2}{\omega_{01}}.
\end{equation}

The calculated loss indicates that flux noise is one of the limiting relaxation mechanisms in our circuit in the vicinity of half flux quantum (red lines on Fig.~\ref{fig:fig6}c). 

\subsubsection{Relaxation due to quasiparticle tunneling}

When quasiparticles from broken Cooper pairs tunnel through one of the Josephson junctions, they can exchange energy with the qubit, leading to relaxation~\cite{PhysRevLett.106.077002, PhysRevB.84.064517}. For the junction $i$ in the circuit, the qubit operator that couples to the noisy quasiparticle bath is $\hat{\mathcal{O}}_\mathrm{qp,i}=\Phi_0\sin(\hat\varphi_i/2)/\pi$, and the noise spectral density is 
\begin{equation}
    S_{\mathrm{qp,i}}^+(\omega)=2\hbar\omega \mathrm{Re}[Y_\mathrm{qp}^{(i)}(\omega)]\coth\left(\frac{\hbar\omega}{2k_BT}\right).
\end{equation}
Here, the real part of the Josephson junction admittance
is~\cite{smith_superconducting_2020} 
\begin{equation}
\begin{split}
    \mathrm{Re}[Y^{(i)}_\mathrm{qp}(\omega)] &= \frac{8e^2}{\Delta}\frac{E_J^{(i)}}{h}\left(\frac{2\Delta}{\hbar\omega}\right)^{3/2}x_\mathrm{qp} \times \\
    &\times \sqrt{\frac{\hbar\omega}{\pi k_BT}} K_0\left(\frac{\hbar|\omega|}{2k_BT}\right) \sinh{\left(\frac{\hbar\omega}{2k_BT}\right)},
\end{split}
    \label{eq:Y_qp}
\end{equation}
where $K_0$ is the modified Bessel function of the second kind, $x_\mathrm{qp}$ is the quasiparticle density, and $\Delta\approx200\,\mu$eV is the superconducting gap of the aluminum. Thus, the relaxation rate is 
\begin{equation}
\begin{split}
    \Gamma_1^\mathrm{qp,i} &= \left|\langle 0|\sin\frac{\hat\varphi_i}{2}|1\rangle\right|^2 \frac{32E_J^{(i)}}{h}x_\mathrm{qp} \times \\
    &\times \sqrt{\frac{2\Delta}{\pi k_BT}} K_0\left(\frac{\hbar\omega_{01}}{2k_BT}\right) \cosh{\left(\frac{\hbar\omega_{01}}{2k_BT}\right)}.
\end{split}
    \label{eq:Y_qp}
\end{equation}
We note that this formula simplifies in the high-temperature limit ($\hbar\omega\gg k_BT$) to
\begin{equation}
    \Gamma_1^\mathrm{qp,i}\approx \left|\langle 0|\sin\frac{\hat\varphi_i}{2}|1\rangle\right|^2 \frac{16E_J^{(i)}}{h}\sqrt{\frac{2\Delta}{\hbar\omega_{01}}}x_\mathrm{qp}.
\end{equation}
In this work, however, we use the full temperature-dependent formula in Eq.~\ref{eq:Y_qp} to predict the relaxation rate, given that the rhombus has a low transition frequency.

In the rhombi qubit, quasiparticle tunneling can occur across any of the four junctions; thus, the total loss due to quasiparticle tunneling is approximately the sum of the contributions of tunneling events across all the junctions if we ignore correlations. In addition, when we associate the external flux with one of the junctions, we can neglect the loss contribution of that particular junction close to frustration~\cite{pop_coherent_2014}. The gold line in Fig.~\ref{fig:fig6}c displays the calculated contribution of quasiparticle tunneling for $x_\mathrm{qp}=10^{-8}$, which shows a similar trend to the loss predicted due to flux noise~\cite{Larson2026}. Thus, both flux noise and quasiparticle tunneling effects are responsible for the reduced relaxation times at the sweet spot in our device.

\end{document}